\documentclass[twocolumn,nofootinbib]{revtex4}%
\usepackage{amsfonts}
\usepackage{amsmath}
\usepackage{amssymb}
\usepackage{graphicx}
\usepackage{dcolumn}
\usepackage{color}

\begin{document} 
\title{Finite-gap twists of carbon nanotubes and an emergent hidden supersymmetry}

\author{Francisco Correa$^{1}$, V\'{\i}t Jakubsk\'y$^{2}$ \\
{\small \textit{$^{1}$Centro de Estudios Cient\'{\i}ficos (CECs), Avenida~Arturo Prat 514, Valdivia, Chile}}\\
{\small \textit{${}^2$Nuclear Physics Institute, \v Re\v z near Prague, 25068, Czech Republic}}}

%
\pacs{11.30.Pb,73.63.Fg,11.30.Na,11.10.Kk}

\begin{abstract}
We consider radially twisted nanotubes in the low-energy approximation where the dynamics is governed by a one-dimensional Dirac equation. The mechanical deformation of the nanotubes is reflected by the presence of an effective vector potential. We discuss twisted carbon and boron-nitride nanotubes, where deformations give rise to periodic and nonperiodic finite-gap Hamiltonians. The intimate relation of these systems with the integrable Ablowitz-Kaup-Newell-Segur hierarchy is exploited in the study of their spectral properties as well as in the computation of the (local) density of states. We show that a nonlinear hidden supersymmetry generated by local supercharges arises naturally in the finite-gap configurations of twisted nanotubes with time-reversal symmetry. The properties of the van Hove singularities are encoded in its structure.

\end{abstract}
\maketitle
\section{Introduction} 
Since their discovery in 1991 \cite{NanotubesObserved}, carbon nanotubes have attracted massive attention from both experimental and theoretical physicists. Their remarkable mechanical and electronic properties, extreme mechanical strength \cite{UltimateStrength}, and elasticity \cite{Elasticity} as well as conductivity \cite{experimentnanotuba} make their use in future electronic devices very promising \cite{NanotubeElectronics}. Carbon nanotubes are also relevant as a low-dimensional test field of fundamental physical theories. For instance, the Klein tunneling, a well known phenomenon in relativistic quantum theory \cite{Klein}, has not been observed for elementary particles so far. However, it was predicted and observed in carbon nanotubes where it stays behind the absence of backscattering on impurities \cite{Ando}.

The single-wall carbon nanotubes are small cylinders with atom-thick shells, that can be created by rolling up graphene nanoribbons \cite{RollTheRibbon}. Despite their structural simplicity (as quasi one-dimensional objects), the nanotubes exhibit a remarkable variety of electronic properties. They can be either metallic or semiconducting, depending on the orientation of the hexagonal lattice in the nanotube. The gap between the valence and conduction band can be regulated by either external fields \cite{ABoscillations} or by mechanical deformations \cite{KaneMele}. This paves the way to strain engineering \cite{StrainEngineering} where  graphene-based devices would be fine-tuned by deformations.

In the present paper, we will consider a class of exactly solvable models of single-wall carbon nanotubes subject to radial twist (axial torsion).  The systems will be studied in the low-energy regime where the dynamics of charge carriers is well approximated by a $(1+1)$-dimensional Dirac equation \cite{Semenoff}.  In particular, the twist configurations under investigation will be described by a finite-gap Hamiltonian.

Besides the analysis of the spectral properties, the computation of the local density of states (LDOS) of the systems will be addressed. This quantity is measurable by means of scanning tunneling microscopy experiments \cite{STM} and is very important for the specification of the electronic properties of nanostructures. When integrated spatially, it provides the density of states (DOS) that reflects the probability of inserting an electron at given energy into the system. 

In general, it is a rather complicated task to analytically compute the local density of states. However, as it was suggested in \cite{Dunne} and demonstrated later in \cite{BdG}, the computation can be considerably simplified for a broad class of Dirac and Bogoliubov - de Gennes Hamiltonians that are characterized by a finite number of gaps in the spectrum. This approach is based on the   intrinsic properties of the finite-gap systems that arise from the stationary AKNS (Ablowitz-Kaup-Newell-Segur) hierarchies of integrable systems \cite{Gesztesy}.

In the next section, we will set up the theoretical framework in which the twisted nanotubes will be analyzed. It will be explained how the Dirac Hamiltonian emerges in the description of twisted carbon (and boron-nitride) nanotubes. In Sec. \ref{AKNSLDOS}, the relevant points of the construction of the AKNS hierarchies are summarized and the closed formula for the LDOS is discussed. Section \ref{examples} is devoted to the presentation of explicit examples {of carbon and boron-nitride nanotubes} where the LDOS and DOS will be computed explicitly. We will show in Sec. \ref{susy} that the singular points of DOS, so-called van Hove singularities, are closely related to the nonlinear supersymmetry that arises naturally for the nanotubes with finite-gap twists and conserved time-reversal symmetry. The last section is left for the discussion and outlook.

\section{The model}\label{model}

Carbon atoms have four valence electrons; three of them are tightly bound in the interatomic bonds while the fourth one is free and can contribute to the electronic properties of the crystal. 
The properties of the collective excitations of these electrons in graphene are well described by the tight-binding Hamiltonian \cite{CastroNeto}. The interactions between the nearest neighbors are only assumed, being specified by a constant hopping parameter. The spin degree of freedom of electrons can be neglected; it is irrelevant in the considered interactions. 

The analysis of the band structure displays the specific feature of graphene: the Fermi surface is formed by six discrete points where valence and conduction bands meet. They are located at the corners of the hexagonal first Brillouin zone  and are called Dirac points.  Only two of them are inequivalent\footnote{The remaining four Dirac points can be reached by translational vectors of the reciprocal lattice, and, hence, do not represent distinct electronic states.} and correspond to different electronic states. Let us denote them as $\mathbf{K}$ and $\mathbf{K'}\equiv-\mathbf{K}$. In the close vicinity of the Dirac points, the energy surface acquires conelike shape. It suggests that the dispersion relation is linear in this region. Indeed, taking $\mathbf{k}=\pm\mathbf{K}+\mathbf{\delta k}$ with $\mathbf{\delta k}\sim 0$ and expanding the tight-binding Hamiltonian up to the terms linear in $\mathbf{\delta k}$, we get the stationary equation for the two-dimensional massless Dirac particle \cite{Semenoff}. The Hamiltonian acquires the same form in both valleys of $\mathbf{K}$ and $\mathbf{K'}$. In the coordinate representation, we have  \footnote{ We set $\hbar=e=1$. $\lambda=\frac{E}{v_F}$ where $E$ is energy and $v_F$ the Fermi velocity of the quasi-particle. }  
\begin{equation}\label{singlevalleyhd}
h(\pm \mathbf{K})\psi_{\pm\mathbf{K}}=(-i\sigma_2\partial_{x}+ i\sigma_1\partial_{y})\psi_{\pm\mathbf{K}}=\lambda\psi_{\pm\mathbf{K}} \, .
\end{equation}
The spinorial degree of freedom in (\ref{singlevalleyhd}), the pseudospin, arises due to the two carbon atoms in the elementary cell; the hexagonal lattice can be thought of as assembled from two triangular lattices. The spin-up or down components of the wave functions are non-vanishing only on one of the two triangular sublattices. The operator $h(\mathbf{K})$ acts on the spinors $(\psi_{\mathbf{K}A},\psi_{\mathbf{K}B})^t$, while $h(\mathbf{K'})$ on the spinors  $(\psi_{\mathbf{K'}B},\psi_{\mathbf{K'}A})^t$.  Here, the first index denotes the valley, the second distinguishes between the sublattices $A$ and $B$ and $t$ denotes transposition.

The formula (\ref{singlevalleyhd}) was introduced already in 1984 by Semenoff \cite{Semenoff} and makes the basis for considering the condensed matter system as a convenient test field for a low-dimensional quantum field theory. Indeed, it makes it possible to observe phenomena in this condensed matter system that are native in high-energy quantum physics, see e.g. \cite{Shytov}, \cite{Gusynin}. 

A single-wall carbon nanotube is rolled up from a straight graphene strip. The actual orientation of the hexagonal lattice in the strip is uniquely determined by the chiral (circumference) vector $\mathbf{C_h}$, which is a linear combination of the translation vectors of the lattice \cite{Blase}. Its length corresponds to the diameter of the nanotube. We can fix the coordinates such that $y$ goes in the circumference direction. Then the chiral vector gets the simple form, $\mathbf{C_h}=(0,C_h)$. 

The effect of rolling up the strip is reflected by the periodic boundary condition imposed on the wave functions, $\psi_{\mathbf{K}}(x,y+C_h)=\psi_{\mathbf{K}}(x,y)$. It leads to the quantization of the momentum in the circumference direction which acquires discrete values $k_y$. In the low-energy approximation, only the value of $k_y$ that minimizes the energy is relevant. The system is then governed by a truly one-dimensional Hamiltonian $-i\sigma_2\partial_{x}+\sigma_1k_y$. 
The actual value of this fixed $k_y$ depends on the character of the nanotube. Instead of going into more details that can be found, for instance in Ref. \cite{Blase}, let us notice that $k_y=0$ corresponds to metallic nanotubes as there is no gap in the spectrum.
When  $k_y\neq 0$, there is a small gap in the spectrum and the nanotube is semiconducting. For purposes of our current analysis, we can suppose that the nanotubes are metallic (i.e. the angular momentum is vanishing, $k_y=0$) and are infinitely long. The latter approximation is rather reasonable due to the recent experiments where ultralong single-wall nanotubes were created \cite{ultralong}. 

Up to now, we considered systems where neither external fields nor any strains were present. By deforming the crystal mechanically, the  interatomic distances in the lattice are modified. Thus, the hopping parameter ceases to be constant  and becomes position dependent. This leads to the appearance of gauge fields in the tight-binding Hamiltonian. It can be approximated in the low-energy limit by Dirac operator with nonvanishing vector potential \cite{KaneMele,Kolesnikov,GaugeFields}. 

We shall consider radial twist of the nanotubes. Let us mention that both single-wall and multi-wall carbon nanotubes with radial twist were used in construction of nanoelectromechanical devices, e.g. single-molecule torsional pendulum \cite{pendulum}, abacus-type resonators \cite{resonators} or even rotors \cite{rotors}. In these systems, the nanotube served as the torsional string that was twisted by deflection of small paddles attached to it; see also \cite{TwistedNanotubes} for a brief review.

The radial twist shifts the atoms in the lattice perpendicularly to the axis, preserving the tubular shape of the nanotube.
The displacement is reflected by a deformation vector, which measures the difference between actual and equilibrium position of atoms. It can be written in our specific case as $\mathbf{d}=(0,d_y(x))^t$. We consider the situation where the displacement is smooth and small on the scale of the interatomic distance. Then the interaction does not mix the valleys of $\mathbf{K}$ and $\mathbf{K'}$, and the system can be studied in the vicinity of one Dirac point only.
The stationary equation for low energy Dirac fermions in the $\mathbf{K}$-valley acquires the following simple form \cite{KaneMele}, \cite{twisting},
\begin{equation}\label{eq1}
 h(\mathbf{K})\phi=(-i\sigma_2\partial_x+\Sigma (x)\sigma_1)\phi=\lambda\phi \, ,
\end{equation}
where the vector potential $\Sigma (x)$ reflects the twist. It is related to the displacement vector by $\mathbf{d}=\zeta\,(0,\int \Sigma (x) dx)$, where $\zeta$ is a constant dependent on the crystal characteristics.\footnote{ $\zeta=\left(-\frac{t}{a}\frac{\partial \ln t}{\partial \ln a}\right)^{-1}$ where $t$ is the hopping parameter and $a$ is the lattice constant. See  \cite{GaugeFields} for more details.}  In this framework, the constant vector potential $\Sigma (x)=\gamma>0$ would reflect a linear displacement ${\mathbf d}=\zeta\,(0,\gamma x)$.

Finally, let us consider the following generalization of (\ref{eq1}), where a mass term is included,
\begin{equation}\label{eq1m}
 h(\mathbf{K})\phi=( -i\sigma_2\partial_x+\Sigma(x)\sigma_1+M\sigma_3)\phi=\lambda\phi \, .
\end{equation}
The analogue of (\ref{singlevalleyhd}) with the mass term $M$ was proposed by Semenoff for description of the quasiparticles in the boron-nitride crystal in the low-energy approximation \cite{Semenoff}. 
The boron-nitride crystal has the same hexagonal structure as graphene. However, the atoms in the elementary cell of the crystal cease to be equivalent. It gives rise to the potential term with $\sigma_3$ that distinguishes between the two triangular sublattices $A$ and $B$. 

We will consider (\ref{eq1m}) as the Hamiltonian of radially twisted carbon ($M=0$) or boron-nitride ($ M\neq 0$), depending on the value of $M$, nanotubes. Boron-nitride nanotubes were studied theoretically and observed experimentally, see e.g. \cite{PreparationBNT}, \cite{TheoryBNT}. Contrary to the carbon nanotubes, they are always semiconducting.

We shall consider the scenario where the term $\Sigma(x)\sigma_1+M\sigma_3$ in (\ref{eq1m}) belongs to the broad class of the finite-gap potentials. In the next section, we will show how the peculiar properties of finite-gap systems can be utilized for computation of the local density of states.
 
\section{Finite-gap twists and the LDOS via AKNS hierarchy}\label{AKNSLDOS}
We review here some properties of the integrable ANKS hierarchies associated with the Dirac Hamiltonian (\ref{eq1m}). To explain more easily the main features, we use the following unitarily transformed Hamiltonian,
\begin{eqnarray}\nonumber
 \tilde{h}&=&\exp\left(-i\frac{\sigma_1\pi}{4}\right)h(\mathbf{K})\,\exp\left(i\frac{\sigma_1\pi}{4}\right)\\&=&\left(\begin{array}{cc}-i\partial_x&\Delta(x)\\\Delta(x)^*&i\partial_x\end{array}\right)\, ,
 \label{h1}
\end{eqnarray}
where $\Delta=\Sigma(x)+i M$. 
This form with diagonal derivative term, the Bogoliubov-de Gennes type Hamiltonian, is used frequently in the analysis of Gross-Neveu and Nambu-Jona-Lasinio models  \cite{Dunne}, \cite{Thies} and will make the presentation more coherent with the specialized literature \cite{Gesztesy}.

The vector potential $\Delta(x)=\Delta$ in (\ref{h1}) is called finite gap (or algebro-geometric in the mathematical literature) when it solves one of the equations from the stationary AKNS hierarchy of the nonlinear differential equations, namely AKNS$_N$. One of the most intriguing properties of the Hamiltonian (\ref{h1}) with a finite gap potential is manifested in its spectrum; it consists of a finite number of bands \cite{Gesztesy}, \cite{concini}. The actual number of bands (or gaps) is fixed by the AKNS$_N$ equation solved by $\Delta$. The values of band-edge energies of a finite-gap system can be obtained in purely algebraic manner; see \cite{Gesztesy}. These features are intimately related with the existence of an integral of motion of the Hamiltonian (\ref{h1}). Nonperiodic finite-gap systems can be obtained as the infinite-period limit of the periodic ones. In this context, the nonperiodic systems are known as kink or kink antikink models in analogy with the soliton solutions in the Korteweg-de Vries (KdV) hierarchy. 

Another relevant feature of this class of models is that they can approximate very well any condensed matter systems described, in the low-energy approximation, by the Hamiltonian (\ref{eq1}) with a generic periodic potential. The Hamiltonian with a generic periodic potential has an infinite number of spectral gaps, the width of which {decreases} rapidly with the increasing absolute value of energy. Hence, the spectrum of such system can be fitted well by a finite- gap one.

The stationary AKNS hierarchy of nonlinear differential equations can be constructed in terms of a Lax pair which consists of the Hamiltonian $\tilde{h}$ and a matrix differential operator $\tilde{S}_{N+1}$, defined as
\begin{equation}\label{S}
 \tilde{S}_{N+1}=i\sum_{l=0}^{N+1}\left(\begin{array}{cc}g_{N+1-l}&f_{N-l}\\f^*_{N-l}&g_{N+1-l}\end{array}\right)\sigma_3h^l \, ,\quad N\in\mathbb{N} \, .
\end{equation}
The functions ${f}_{n}(x)$ and ${g}_n(x) $ are defined, recursively, in the following manner,
\begin{eqnarray}
    {f}_n&=&-\frac{i}{2}{f}_{n-1}^{\,\prime}+\Delta \,
    {g}_n \, ,\\
    {g}_n^\prime &=& i\left(\Delta^*{f}_{n-1}-\Delta\, {f}_{n-1}^*\label{int}
    \right)\, ,\\
    {g}_0&=&1,\quad
    {f}_{-1}=0 \, .
    \label{recursion}
\end{eqnarray}
The functions $f_{n}$ and $g_n$ depend on $\Delta(x)$ and its derivatives and also contain integration constants that appear due to the integration of (\ref{int}); see Ref. \cite{BdG} for details.

The operator (\ref{S}) satisfies the following commutation relation  for any positive integer $N$,
\begin{eqnarray}\label{commutator}
 [\tilde{S}_{N+1},\tilde{h}]&=&2i\left(\begin{array}{cc}0&f_{N+1}\\-f^*_{N+1}&0
                   \end{array}
\right) \, ,
\end{eqnarray}
The stationary AKNS hierarchy of nonlinear differential equations is then defined in terms of the vanishing commutator (\ref{commutator}),
\begin{equation}\label{AKNS1}
\text{AKNS}_N=f_{N+1}=0 \, .
\end{equation}
The Hamiltonian  $\tilde{h}$ and the operator $\tilde{S}_{N+1}$ are called the Lax pair of the stationary AKNS hierarchy.  When a function $\Delta$ satisfies the $(N+1)^{\text{th}}$ order differential equation AKNS$_{N}$, all the next equations of the hierarchy with greater values than $N$ (and with the integration constants fixed appropriately\footnote{The AKNS$_{N}$ can be written as a linear combination $\sum_{l=0}^{N+1}c_{l}\hat{{f}}_{l}$ where the functions $\hat{{f}}_{l}$ are defined like in (\ref{recursion}) but with all the integration constants that emerge in (\ref{int}) fixed to zero.}) are immediately solved.   

The operators $\tilde{h}$ and $\tilde{S}_{N+1}$ satisfy  the remarkable algebraic relation,
\begin{equation}\label{BuchanalChaudy}
 \tilde{S}_{N+1}^2=\prod_{n=0}^{2N+1}(\tilde{h}-\lambda_{n}) \, ,
\end{equation}
where $\lambda_{n}$ are band-edge energies. The operator valued function on the righthand side is known as the spectral polynomial. 
The integral $\tilde{S}_{N+1}$ annihilates all the singlet eigenstates $\tilde{\phi}_n$ of $\tilde{h}$, corresponding to the band-edge energies, $(\tilde{h}-\lambda_n)\tilde{\phi}_n=0$,
\begin{equation}\label{Skernel}
\tilde{S}_{N+1} \tilde{\phi}_n=0, \quad n=0,1,...,2N+1 \, .
\end{equation}

The local density of states $\rho(x,\lambda)$ is defined in terms of the trace of the Green's function, $\tilde{R}(x,\lambda)\equiv \tilde{G}(x,x,\lambda)$,
\begin{eqnarray}\label{LDOS}
  \rho(x,\lambda)&=&-\frac{1}{\pi}\lim_{\mbox{Im}\lambda\rightarrow 0_+}\mbox{Im}\,Tr \,\tilde{R}(x,\lambda)\,,
\end{eqnarray}
where the trace is computed over matrix degrees of freedom.
The function $\tilde{R}(x,\lambda)$ is also called diagonal resolvent or Gorkov resolvent.

The spatial integration of LDOS leads to the formula for DOS. In case of periodic quantum systems, the integration can be performed over one period $L$ \footnote{For the nonperiodic settings, the spatial integration can be divergent. }, 
\begin{equation}\label{dos}
\text{DOS}=\frac{1}{L}\int_{L} \rho(x,\lambda)\, dx\,.
\end{equation}

Explicit calculation of the Green's function  can be quite difficult. 
Nevertheless, the definition (\ref{LDOS}) suggests that the need of its explicit knowledge might be avoided; only the diagonal resolvent is required to find LDOS. This fact was utilized in \cite{Dunne} and further developed in \cite{BdG}. Indeed, an exact form of the diagonal resolvent was found for a wide class of Hamiltonians (\ref{h1}). The approach was based on the fact that $\tilde{R}(x,\lambda)$ has to satisfy the Dikii-Eilenberger equation \cite{eilenberger},
\begin{eqnarray}
\frac{\partial}{\partial x}\tilde{R}(x; \lambda)\, \sigma_3-i\, \left[
\begin{pmatrix}
\lambda &-\Delta(x) \cr
\Delta^*(x) & -\lambda
\end{pmatrix}, \tilde{R}(x; \lambda)\,\sigma_3
\right]
&=&0\, ,\nonumber\\ 
\label{dikii}
\end{eqnarray}
where $\lambda$ belongs to the spectrum of $\tilde{h}$. Additionally, $\tilde{R}(x,\lambda)$ has to satisfy the following requirements,
\begin{equation}\label{constrains}
 \tilde{R}=\tilde{R}^{\dagger},\quad \det \tilde{R}=-\frac{1}{4}\, , 
\end{equation}
where the latter one fixes the normalization of $\tilde{R}$. For more details on the properties of $\tilde{R}$ and derivation of (\ref{constrains}), see e.g. the Appendix in \cite{kos}.

Making the following ansatz for the diagonal resolvent \cite{BdG},
\begin{eqnarray}
\tilde{R}(x; \lambda)=\sum_{n=0}^{N+1} \beta_n(\lambda)\, \begin{pmatrix}
    {g}_n(x) & {f}_{n-1}(x)\cr
    {f}_{n-1}^*(x) & {g}_n(x)
    \end{pmatrix}\, ,
    \label{poly}
    \end{eqnarray}
and substituting (\ref{poly}) into (\ref{dikii}), the Dikii-Eilenberger equation transforms into the two (mutually conjugated) nonlinear differentials equations of the form of the AKNS hierarchy.  The diagonal entries in (\ref{dikii}) vanish identically due to the recurrence relations (\ref{recursion}). The resulting equation can be written as 
\begin{eqnarray}\label{AKNS}
 &&\sum_{n=0}^{N+1}\beta_n(\lambda){f}_n-\lambda\sum_{n=0}^{N+1}\beta_n(\lambda){f}_{n-1}=0 \, ,
\end{eqnarray}
which can be solved by fixing properly the constants $\beta_n(\lambda)$, see footnote $4$ and Ref. \cite{BdG}. It can be shown that the ansatz (\ref{poly}) fulfills the requirements (\ref{constrains}). \footnote{ The ansatz (\ref{poly}) is manifestly hermitian. Additionally, it also satisfies the second condition in (\ref{constrains}). Indeed, one can check directly that $(\mbox{det}\tilde{R}(x,\lambda))'=0$ with the use of (\ref{AKNS}). }

Making the inverse transformation (\ref{h1}), we can find the Lax operator associated with the finite-gap Hamiltonian (\ref{eq1m}) as
\begin{align}\label{Ss} 
&{S}_{N+1}=\exp\left(i\frac{\sigma_1\pi}{4}\right)\tilde{S}_{N+1}\exp\left(-i\frac{\sigma_1\pi}{4}\right)& \\
&=-i\sum_{l=0}^{N+1}\left(g_{N+1-l}\mathbf{1}+\sigma_3\mbox{Im}f_{N-l}+\sigma_1\mbox{Re}f_{N-l}\right)\sigma_2h^l& \, .\notag
\end{align}
The diagonal resolvent for the Hamiltonian $h$ can be obtained directly from (\ref{eq1m}), since the trace of an operator is invariant with respect to similarity transformations.

\section{Exactly solvable models of the twisted nanotubes}\label{examples}

The periodic systems described by (\ref{eq1}) can be classified in terms of a quantity which we call average twist. It is defined as
\begin{equation}\label{constant}
 \Sigma_c =\frac{\mbox{max}(\Sigma)+\mbox{min}(\Sigma)}{2},
\end{equation} 
and corresponds to the value around which the potential is oscillating. 
We will present two- and four-gap systems, denoted as $\Sigma(x)=\Delta_2$ and $\Sigma(x)=\Delta_4$, respectively, where the average twist is vanishing. Then we will consider two simple cases where it acquires nonzero values. They correspond to the one-, $\Sigma(x)=\Delta_1$, and three-gap  $\Sigma(x)=\Delta_3$ systems. The mass term will be identically zero in all these models, $M=0$. We will see that the actual value of the average twist is in correlation with the qualitative spectral properties of  these models. 

As the last example, we will consider a nonperiodic system with a constant mass, $M\neq0$. It will serve for illustration of a twisted boron-nitride nanotube.

\subsection{Configurations with zero average twist}

\subsubsection{Two-gap system}
\begin{figure}
\centering
\begin{tabular}{cc}
 \includegraphics[scale=0.675]{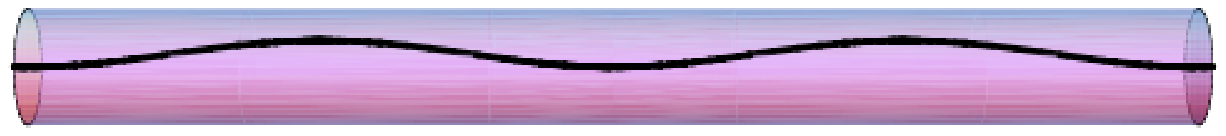}\\ \\
   \includegraphics[scale=0.675]{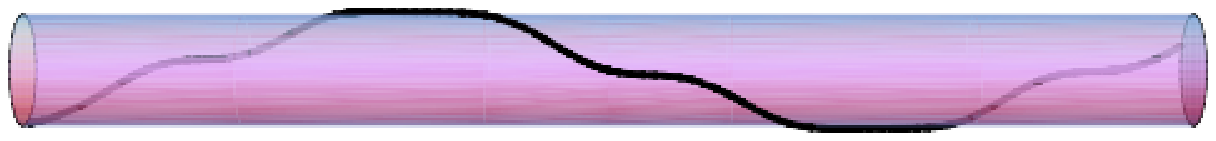}\\ \\ 
  \includegraphics[scale=0.675]{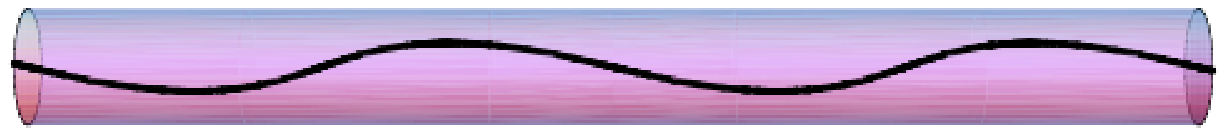}
\end{tabular}
\caption{Illustration of the two-gap (upper), three-gap (middle)  and four-gap (lower) configurations of the twisted carbon nanotubes. In the system without twist, the black line would be straight. } 
\label{twist}
\end{figure}

First, let us consider the system governed by (\ref{eq1}) with the vector potential 
\begin{equation}\label{crystal1}
\Delta_{2} = m  k^2 \frac{{\rm sn }\,m x \, {\rm cn }\, mx}{{\rm dn }\,m x} \, ,
\end{equation}
where $m$ is a real parameter and $k\in(0,1)$.
This vector potential is induced by the deformation specified by the following displacement vector (see Fig. \ref{twist}),
\begin{equation}
{\bf d}=(0,-\zeta\, \ln {\rm dn }\,m x) \, .
\end{equation}

The crystal kink two-gap potential (\ref{crystal1}) is given in terms of doubly periodic Jacobi elliptic functions depending on the modular parameter $k$. It has a real period $L=2 K(k)$,  where $K(k)$ is the complete  elliptic integral of the first kind. For the definitions and properties of the elliptic functions, we recommend Refs. \cite{Abramovitz, WW}. The infinite-period limit ($k \to 1$) of (\ref{crystal1}) is called the single kink vector potential $\Delta_2= \tanh x$. Let us notice that the properties of the Dirac electron in graphene in the presence of a single-kink-type vector potential were analyzed in \cite{milpas}. The potential $\Delta_{2}$ vanishes in the limit when the modular parameter goes to zero.

The spectrum of the one-dimensional Dirac Hamiltonian $h(\mathbf{K})$ has two gaps located symmetrically with respect to zero. The band-edge energies are $\lambda_0=-\lambda_3=-m$ and $\lambda_1=-\lambda_2=-m \sqrt{1-k^2}$. The corresponding eigenstates ($(h(\mathbf{K})-\lambda_n)\phi_n=0$, $n=0,1,2,3$) are
\begin{align}\label{2gf1}
\phi_{0}=&\left( -{\rm sn }\,m x  ,\,  \frac{{\rm cn }\,m x}{{\rm dn }\,m x}\right)^t\, ,&  \phi_{3}= \sigma_3 \phi_{0} \, ,\\
\phi_{1}=&\left(\frac{1}{\sqrt{1-k^2}} {\rm cn }\,m x  ,\,  \frac{{\rm sn }\,m x}{{\rm dn }\,m x} \right)^t \, ,&  \phi_{2}= \sigma_3 \phi_{1} \, .
\label{2gf2}
\end{align}
The band-edge energies are nondegenerate, while the energies from the interior of the bands are doubly degenerated.

\begin{figure}[h!]
\centering
\includegraphics[scale=0.65]{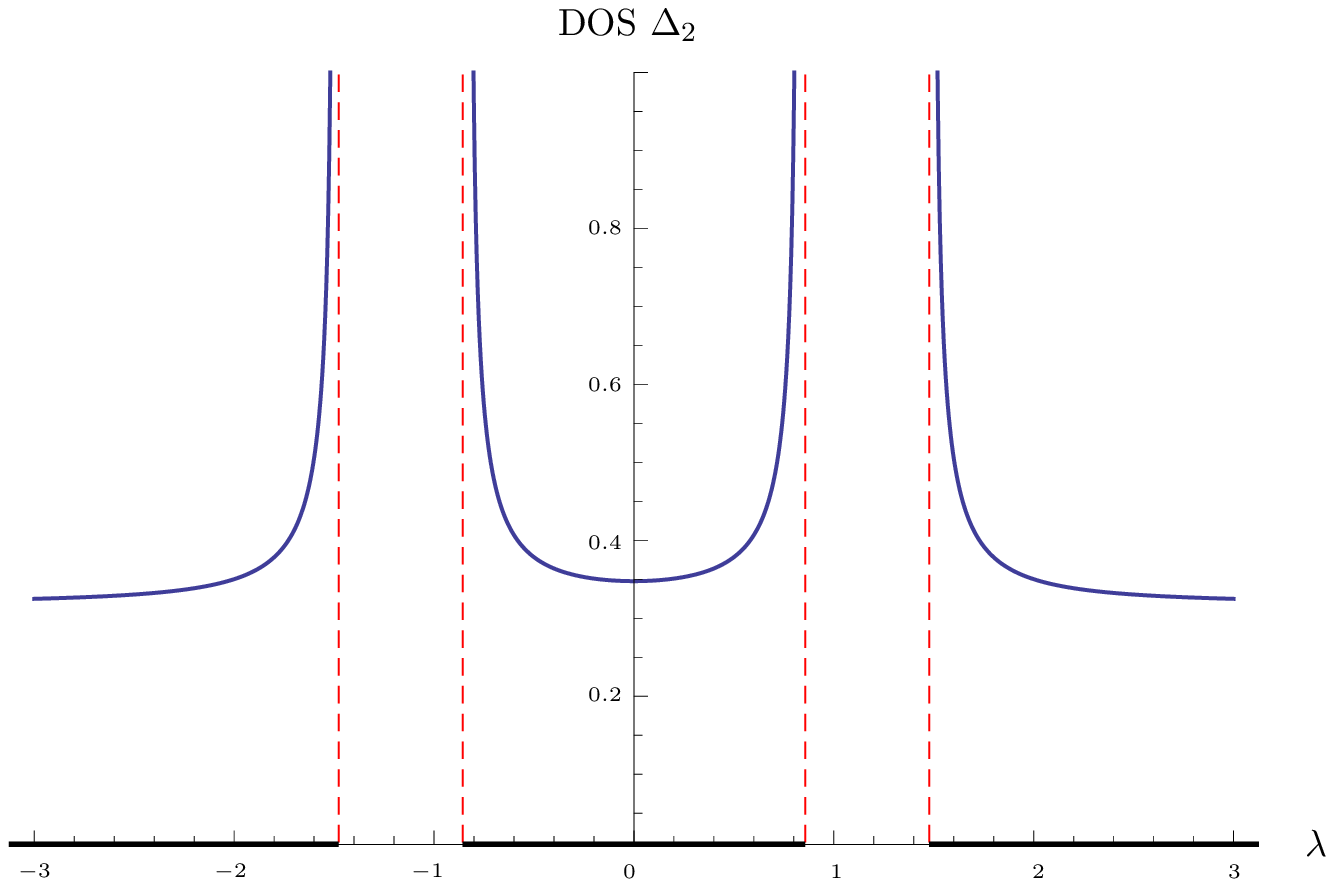}
\caption{Illustration of DOS (\ref{dos2}) of the Hamiltonian (\ref{eq1m})  with  $\Sigma(x)=\Delta_{2}$ and $M=0$ with $m=1.5$ and $k=0.7$.}\label{dostwogap}
\end{figure}

Using directly the formula (\ref{poly}) for $N=2$, we can find the explicit form of the diagonal resolvent. Its trace then reads
\begin{equation}\label{2gaptrace}
{\rm Tr}\, R_2(x; \lambda)= \frac{\lambda^2+\frac{1}{2}\Delta_2^2}{\sqrt{(m^2-\lambda^2)(\lambda^2+m^2(k^2-1))}} \, ,
\end{equation}
and the associated density of states acquires the following form
\begin{equation}\label{dos2}
\text{DOS}_{2}=-\frac{1}{\pi}\lim_{\mbox{Im}\lambda\rightarrow 0_+}\mbox{Im}\,\frac{\lambda^2-m^2\frac{E(k)}{K(k)}}{\sqrt{(m^2-\lambda^2)(\lambda^2+m^2(k^2-1))}} \, ,
\end{equation}
where we have used Eqs. (\ref{LDOS}) and (\ref{dos}). Notice that $\text{DOS}_{2}$ is identically zero when $\lambda$ belongs to the prohibited gaps. The function in the argument is purely real for these values of $\lambda$ and, thus, the imaginary part is vanishing identically, see Fig. \ref{dostwogap}.

\noindent

\subsubsection{Four-gap system}

As the next example, we shall consider the $2 K(k)$-periodic system described by the Hamiltonian (\ref{eq1}) with the vector potential
\begin{equation}\label{crystal4}
\Delta_4 = 6m k^2 \frac{ {\rm sn }\,m x \, {\rm cn }\, mx  \, {\rm dn }\, mx}{1+k^2+\delta-3k^2{\rm sn }^2\,m x }\, ,
\end{equation}
where $\delta=\sqrt{1-k^2+k^4}$. The crystal kink four-gap potential (\ref{crystal4}) is an isospectral deformation of the crystal kink potential $\Delta_{4'} = 2 m k^2 \frac{{\rm sn }\,m x \, {\rm cn }\, mx}{{\rm dn }\,m x}$. Both potentials reduce to the single kink  $\Delta_{4} =\Delta_{4'} =2 m \tanh mx$ when $k\to 1$. The associated displacement vector in this case takes the form
\begin{equation}
{\bf d}=(0,- \zeta\,\ln (1+k^2+\delta-3k^2{\rm sn }^2\,m x))
\end{equation}
and is illustrated in Fig. \ref{twist}.
The spectrum of $h(\mathbf{K})$ has five bands and eight band-edge states $\phi_n$, $n=0,...,7$, which can be defined with help of and an operator ${\cal D}=\frac{d}{dx}+\Delta_4$ as follows, \footnote{This way to express the eigenfunctions is just the essence of usual supersymmetric quantum mechanics applied for finite-gap potentials. To avoid the details here, we refer to \cite{samsonovetal, trisusy}.} 
\begin{equation}\label{phi}
\phi_n= \left(\psi_n,\frac{1}{\lambda_n}{\cal D}\psi_n\right)^t,\quad (h(\mathbf{K})-\lambda_n)\phi_n=0.
\end{equation}
Keeping in the mind the spectral symmetry $\lambda\leftrightarrow -\lambda$ (which is valid for any  model (\ref{eq1m}) with $M=0$), it is sufficient to  find just the first four  eigenstates $\phi_0,\ \phi_1,\ \phi_2,\ \phi_3$, since the remaining four can be obtained as $\phi_{n+4}=\sigma_3\phi_n$, where $n=0,1,2,3$. They are given in terms of the following functions
\begin{flalign}
&\psi_0=m\left(1+k^2-\delta-3k^2{\rm sn}^2\, mx \right), \quad \lambda_0=-2m\sqrt{\delta} \, ,&\\ 
&\psi_1={\rm cn}\, x\, {\rm sn}\, x  , \quad
\lambda_1=-m\sqrt{2-k^2+2\delta} \, ,& \\
&\psi_2  ={\rm dn}\, x\, {\rm sn}\, x , \quad
\lambda_2=-m\sqrt{2k^2-1+2\delta} \, , &\\
&\psi_3= {\rm cn}\, x\, {\rm dn}\, x , \quad \lambda_3=-m\sqrt{2\delta-1-k^2}  \, ,&
\end{flalign}
where $\lambda_n$ are the corresponding eigenvalues. 

\begin{figure}[h!]
\centering
\includegraphics[scale=0.65]{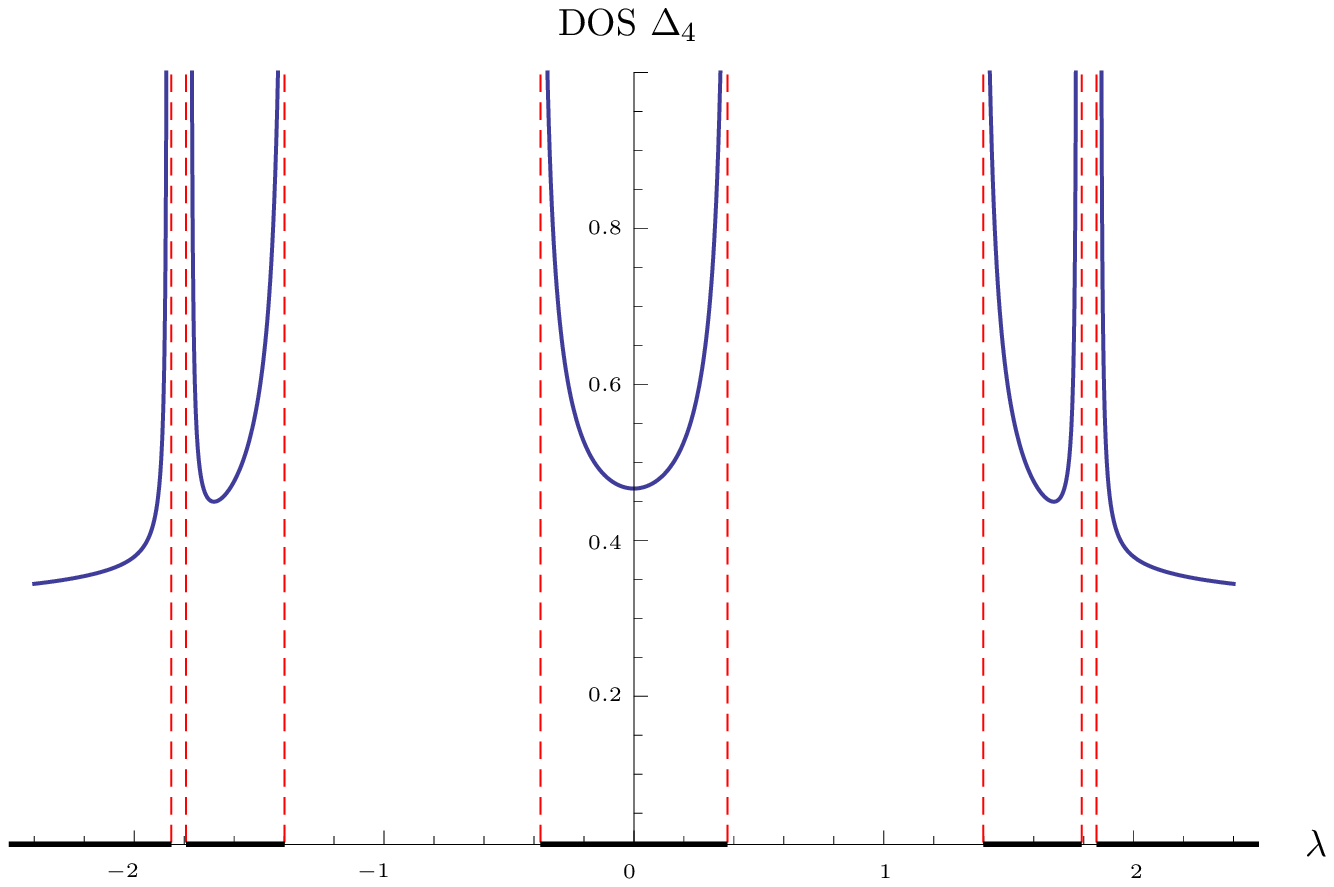}
\caption{Illustration of DOS of the Hamiltonian (\ref{eq1m})  with $\Sigma(x)=\Delta_{4}$ and  $M=0$ with $m=1$ and $k=0.6$.}\label{dosfourgap}
\end{figure}

The local density of states can be computed using the method described in the preceding section,
\begin{equation} \label{4gaptrace}
{\rm Tr}\, R_4(x; \lambda)=\frac{\alpha_1+\alpha_2\Delta_4^2 +3\Delta_4^4+\Delta_4'^2-2\Delta_4 \Delta_4''}{8\sqrt{(\lambda_0-\lambda^2)(\lambda^2 -\lambda_1^2)(\lambda^2 -\lambda_2^2)(\lambda^2 -\lambda_3^2)}},
\end{equation}
where the constants are
\begin{eqnarray}
\alpha_1 &=& 8(\lambda^4-5m^2\delta\lambda^2+4m^4\delta), \\
\alpha_2 &=& 4(\delta^2-5m^2\delta^2)  \, .
\end{eqnarray}
The explicit (analytical) form of the density of states is rather cumbersome. In Fig.  \ref{dosfourgap}, we present the numerical computed DOS of the current four-gap system.

Let us note that for the system described by the isospectral  potential $\Delta_{4'}$, the resolvent trace 
\begin{equation} \label{4gaptraceb}
{\rm Tr}\, R_{4'}(x; \lambda)=\frac{\eta_1+\eta_2\Delta_{4'}^2+\eta_3  \Delta_{4'}^4}{8\sqrt{(\lambda_0-\lambda^2)(\lambda^2 -\lambda_1^2)(\lambda^2 -\lambda_2^2)(\lambda^2 -\lambda_3^2)}}
\end{equation}
can be written just in terms of a polynomial in $\Delta_{4'}$, where $\eta_1$, $\eta_2$ and $\eta_1$ are constant depending on $\lambda$.

\subsection{Semiconducting carbon nanotubes via nonzero average twist}

\subsubsection{One-gap system}
The simplest example of a finite-gap system with the nonzero average twist is given by the Hamiltonian (\ref{eq1}) with the constant vector potential $\Sigma(x)=\gamma$. The two band-edge energies correspond to $\lambda_0=-\lambda_1=-\gamma$. The local density of states can be found in the following form
\begin{equation}\label{0gaptrace}
{\rm Tr}\, R_4=\frac{\lambda}{\sqrt{\gamma^2-\lambda^2}}\, .
\end{equation}
The constant potential can be regarded as periodic with the period being equal to any real number $L$. We can compute the average twist as $\Sigma_c=\gamma$. 

The spectrum of the system has two
bands separated by a gap of width $2\Sigma_c$. This suggests that the central gap
is twice the average twist.

\subsubsection{Three-gap system}
Let us test the suggestion in the case of a more complicated system. Its Hamiltonian (\ref{eq1}) has the  $2K(k)$-periodic vector potential 
\begin{equation}\label{crystal2}
    \Delta_3=\frac{{\rm cn}\,b\, {\rm dn}\,b}{{\rm sn}\,b}+k^2\, {\rm sn}\,b\,{\rm sn}
    \,(x) \,{\rm sn}\,\left(x+b\right), 
\end{equation}
which is called the crystal kink-antikink, three-gap potential, \cite{BdG}. The real parameter $b\in(0, K(k))$ represents the distance between the kink and the antikink.

The vector potential is induced by the displacement ${\bf d}=(0,\zeta\,F(x))$, where $F(x)$ is as follows,
\begin{equation}\notag
F(x)\!=\!\frac{{\rm cn}\,b\, {\rm dn}\,b}{{\rm sn}\,b} \, \Pi(k^2 \, {\rm sn}^2 b,{\rm am}\; x |k) -\frac{1}{2}\ln \left(\!1\!-\! k^2 \,  {\rm sn }^2 b  \, {\rm sn }^2 x\right).
\end{equation}
The function $\Pi(a;x | \phi)$ is the incomplete elliptic integral of the third kind and ${\rm am}\; x  $ is the Jacobi amplitude. See Fig. \ref{twist} for illustration.

When $b= K(k)$, (\ref{crystal2}) is reduced to the two-gap vector potential (\ref{crystal1}). In the infinite period limit, the single kink-antikink solution is recovered \cite{Jackiw},  $\lim_{k\rightarrow1}\Delta_1(x)=\coth b +\tanh x -\tanh (x+b)$.

The spectrum of (\ref{eq1}) with (\ref{crystal2}) contains three gaps positioned symmetrically with respect to zero. The three band-edge states with negative energies are
\begin{eqnarray}\notag
\phi_{0}&=&\left(  -{\rm sn }\,(x),\,\,{\rm sn }\,(x+b) \right)^t  \, , \quad \lambda_{0}=-\frac{1}{{\rm sn }\,b} \, ,  \\
\phi_{1}&=&\left(-{\rm cn }\,(x) ,\,\,  {\rm cn }\,(x+b)\right)^t  \,  , \quad \lambda_{1}=-\frac{{\rm dn }\,b}{{\rm sn }\,b} \, , \notag \\
\phi_{2}&=&\left( -{\rm dn }\,(x),\,\, {\rm dn }\,(x+b) \right)^t  \,  , \quad \lambda_{2}=-\frac{{\rm cn }\,b}{{\rm sn }\,b}\, . \label{states3}
\end{eqnarray}
The positive energy states are obtained as  $\phi_{n+3}=\sigma_3\phi_n$ and correspond to the energies $\lambda_{n+3}=-\lambda_n$,  where $n=0,1,2$.

\begin{figure}[h!]
\centering
\includegraphics[scale=0.65]{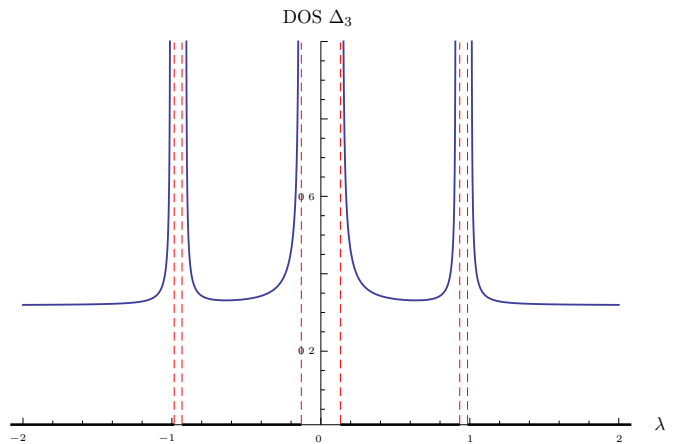}
\caption{Illustration of DOS of the Hamiltonian (\ref{eq1m})  with  $\Sigma(x)=\Delta_{3}$ and $M=0$ with $k=0.2$ and $b=1.5$.} \label{dosthreegap}
\end{figure}

The trace of diagonal resolvent can be computed from (\ref{poly})  for $N=2$ in the following form
\begin{equation}\label{3gaptrace}
{\rm Tr}\, R_3(x; \lambda)=\frac{\frac{\lambda}{2}\left(\alpha+\Delta^2_3 \right)}{\sqrt{( \lambda_0^2-\lambda^2)(\lambda_1^2-\lambda^2 )(\lambda_2^2-\lambda^2 )}}\, ,
\end{equation}
where $\alpha=1+k^2+2\lambda^2-\frac{3}{{\rm sn}^2\,b}$. The actual integration of the formula above, needed for analytical form of DOS, is rather complicated. We present Fig. \ref{dosthreegap} of DOS for the three-gap case that was obtained numerically.

The average twist associated with the potential (\ref{crystal2}) can be found as
\begin{equation}
 \Sigma_c=\frac{{\rm cn}\,  b}{{\rm sn}\,  b}.
\end{equation}
Checking the corresponding band-edge energies $\lambda_2$ and $\lambda_3$ in (\ref{states3}), we can see that that the gap between the positive and negative energies is exactly of width $2\Sigma_c$. 

Comparing the spectra of the presented systems, we can see that the nonvanishing average twist (\ref{constant}) is proportional to the magnitude of the central spectral gap in the system.  In the two- and four-gap systems, the average twist is vanishing and there is no gap between positive and negative energies. These nanotubes are conducting in the sense that infinitesimal excitation is sufficient to kick the electrons from valence band to conduction band. The systems with the nonvanishing average twist are different. They have a gap between positive and negative energies that are equal to $2\Sigma_c$ and, hence, are semiconducting.

\subsection{Boron-nitride nanotubes}
As the last example, we shall consider a \textit{nonperiodic} system with the nontrivial mass term. We take the potential term of (\ref{eq1m}) in the following form,
\begin{equation}\label{complexcrystal}
\Sigma(x)= N \sin \frac{\theta}{2} \tanh(\sin \frac{\theta}{2}x ), \quad M=-N \cos \frac{\theta}{2},
\end{equation}
where $N$ is a positive integer and $\theta$ is a real parameter. 
The potential is classified as $(N+1)$-gap as it solves the corresponding equation of the AKNS. It has $N+2$ singlet states in the spectrum. Two of them correspond to the energies that form the threshold of the continuum spectrum, the rest is associated with bound states of the system.
In the case of $N=1$, the eigenstates are then given as
\begin{align}
&\psi_0=\left( \tanh \left(\sin \frac{\theta}{2}x \right),-\cot \frac{\theta}{4} \right)^t,& \, &\lambda_0=-1,& \\
&\psi_1=\left({\rm sech} \, \left(\sin \frac{\theta}{2}x\right),\,0\right)^t,& \, &\lambda_1=\cos \frac{\theta}{2},& \\
&\psi_2=\left(\tanh \left(\sin \frac{\theta}{2}x \right),\tan \frac{\theta}{4}  \right)^t,& \, &\lambda_2=1.&
\end{align}
The trace of the diagonal resolvent can be computed in the following manner,
\begin{equation} \label{complextrace}
{\rm Tr}\, R(x; \lambda)=\frac{\lambda^2-\lambda \cos \frac{\theta}{2}-\frac{1}{2}\sin^2 \frac{\theta}{2} \, {\rm sech} ^2\, \left(\sin \frac{\theta}{2}x \right) }{\sqrt{1-\lambda^2}(\lambda-\cos \frac{\theta}{2})} \, .
\end{equation}
\vspace{3mm}

It is worth noticing  that in the examples of the carbon nanotubes, the trace of diagonal resolvent (\ref{2gaptrace}), (\ref{4gaptrace}),  (\ref{4gaptraceb}),  and (\ref{3gaptrace}) could be written in terms of the finite-gap potential and its derivatives. A similar result is known for the Schr\"odinger systems with Lam\'e potential. The trace of the diagonal resolvent corresponded in that case to a polynomial of the finite-gap potential \cite{belokolos}. This is related to the fact that  the square of the Dirac operator with the four-gap potential $\Delta_{4'}$ corresponds to an extended Schr\"odinger operator with two-gap Lam\'e potential. 
 
In the current case with the nonvanishing mass term, the trace of diagonal resolvent can be written as a function of the amplitude of the corresponding complex potential $\Delta$ [see (\ref{h1})],  where $|\Delta|^2=\Sigma(x)^2+M^2$. For (\ref{complexcrystal}), we can write
\begin{equation}\nonumber
{\rm Tr}\, R(x; \lambda)=\frac{P_{2N}(x,\theta)}{\sqrt{N^2-\lambda^2}\prod_{n=1}^{2N}(\lambda-\lambda_n)^2}.
\end{equation}
Here,
\begin{equation}\label{complexgeneral}
P_{2N}(x,\theta)=\sum_{n=0}^N c_n(\lambda) \left( \sin^2 \frac{\theta}{2} \, {\rm sech} ^2\, (\sin \frac{\theta}{2}x) \right)^n 
\end{equation} 
with $c_n(\lambda) $ being specific constants \footnote{
For $N=2$ the band edge energies are  $\lambda_0=-\lambda_4=-2, \, \lambda_1=-\lambda_3= -\sqrt{\frac{7+\cos \theta}{2}} $ and $\lambda_2=2\cos \frac{\theta}{2} $. The constants in (\ref{complexgeneral}) are then given in the following form
\begin{align*}
c_0(\lambda)=\lambda ^4-2 \lambda ^3 \cos \left(\frac{\theta }{2}\right)-\frac{1}{2}
   \lambda ^2 (\cos (\theta )+7)+ \\
   \frac{1}{2} \lambda  \left(15 \cos
   \left(\frac{\theta }{2}\right)+\cos \left(\frac{3 \theta
   }{2}\right)\right)\,  \\
   c_1(\lambda)=2 \left(-\lambda ^2+\lambda  \cos \left(\frac{\theta}{2}\right)+\cos (\theta
   )+1\right), \, c_2=\frac{9}{4}
\end{align*}
}.

\section{van Hove singularities and the nonlinear supersymmetry}\label{susy}

The densities of states have a set of singular points that are called van Hove singularities. A closer inspection of the corresponding formulas (\ref{2gaptrace}), (\ref{4gaptrace}), (\ref{0gaptrace}), (\ref{3gaptrace}), and (\ref{complextrace}) shows that the number as well as the position of the van Hove singularities coincide precisely with the singlet band-edge energies of the finite-gap systems.
In this section, we will show that this coincidence is reflected by a nonlinear supersymmetry that underlies the finite-gap configurations of the twisted nanotubes. 

Quantum systems in presence of a magnetic field cease to be time-reversal invariant. The time-reversal operator changes the sign of momentum while it preserves the coordinate. It changes the sign of the magnetic field.  
It can be represented by an anti-unitary operator $\mathcal{T}$ that satisfies $\mathcal{T}^{\dagger}\mathcal{T}=1$, $\mathcal{T}i\mathcal{T}=-i$ and $\mathcal{T}^2=-1$. The latter equality arises due to the half-integer spin of the considered particles.

One can check that the anti-unitary operator  $\sigma_2 T$ ($T$ denotes complex conjugation, $T^2=1$) does not commute with the Hamiltonian (\ref{eq1m}) due to the symmetry breaking term $\Sigma(x)$ (and $M$).
However, we have to keep in mind that these terms arise from the tight-binding model, which, despite the deformations of the crystal, is time-reversal invariant \cite{CastroNeto}.

The time-reversal symmetry of the system in the low-energy regime emerges when dynamics in both valleys corresponding to $\mathbf{K}$ and $\mathbf{K'}$ is taken into account.
The total Hamiltonian reads 
\begin{equation}\label{totalh}
 \mathcal{H}=\left(\begin{array}{cc}h(\mathbf{K})&0\\0&h(-\mathbf{K})\end{array}\right)\, ,
\end{equation}
where the energy operators  $h(\pm\mathbf{K})$ of the subsystems are given as
\begin{equation}\label{htwovalley}
 h(\pm\mathbf{K})=-i\sigma_2\partial_{x}\pm \Sigma(x)\sigma_1 \pm M\sigma_3\, .
\end{equation}
The operator (\ref{totalh}) acts on the bispinors $\Psi=(\psi_{\mathbf{K}A},\psi_{\mathbf{K}B},\psi_{\mathbf{K}'B},\psi_{\mathbf{K}'A})$, where we use the notation introduced in the second section below (\ref{singlevalleyhd}).

The Hamiltonian $\mathcal{H}$ commutes with the time-reversal operator $\mathcal{T}$ which is defined in the following manner\footnote{In \cite{Gusynin}, the real spin of electrons in taken into account. There, the time-reversal operator is defined as $(\sigma_{1}\otimes\sigma_1)T\sigma_2$, where the last Pauli matrix acts on the spin degree of freedom of the electrons. As we do not consider real spin of electrons in our model, we have to define $\mathcal{T}$ as in (\ref{tr}) to keep $\mathcal{T}^2=-1$.},
\begin{equation}\label{tr}
 [\mathcal{H},\mathcal{T}]=0,\quad \mathcal{T}=\sigma_1\otimes\sigma_2\, T.
\end{equation}
As the considered system consists of a single fermion, the Kramer's theorem applies; all the energy levels of (\ref{totalh}) have to be at least doubly degenerate.  In case of a periodic system, the band structure of $h(\mathbf{K})$ is determined by $2N+2$ nondegenerate band-edge energies $\lambda_n$. In the infinite period limit, the operator has $N+2$ singlet states. As we can see from (\ref{htwovalley}), the operators $h(\pm\mathbf{K})$ are unitarily equivalent, $h(\mathbf{K})=\sigma_2h(-\mathbf{K})\sigma_2$.  Hence, $\mathcal{H}$ has the same band structure as $h(\mathbf{K})$, but the degeneracy is doubled as is required by the Kramer's degeneracy theorem.

Degeneracy of energy levels is reflected by a set of integrals of motion that are based on the Lax integral $S_{N+1}$, see (\ref{Ss}). In  the individual subsystems governed by $h(\pm\mathbf{K})$,  the degeneracy is associated with two diagonal operators, $\mathcal{Q}_0$, and $\mathcal{Q}_3$,
\begin{equation}\label{q03}
 \mathcal{Q}_0=\left(\begin{array}{cc}S_{N+1}&0\\0&\sigma_2S_{N+1}\sigma_2\end{array}\right),\quad \mathcal{Q}_3=\tau_{30}\mathcal{Q}_0,
\end{equation} 
where $\tau_{30}=\sigma_3\otimes \mathbf{1}$.
The intervalley (Kramer's) degeneracy is naturally reflected by the operators  $\mathcal{Q}_{1}$ and $\mathcal{Q}_{2}$,
\begin{equation}\label{set3}
 \mathcal{Q}_1=\tau_{12}\mathcal{Q}_0,\quad \mathcal{Q}_2=\tau_{22}\mathcal{Q}_0,
\end{equation}
where $\tau_{ab}=\sigma_a\otimes\sigma_b$, $a,b=1,2$. All these operators commute with the total Hamiltonian,
\begin{equation}\label{set4}
 [\mathcal{Q}_a, \mathcal{H}]=0\, .
\end{equation}
By construction, these operators close Lie algebra $so(3)\oplus u(1)$,
\begin{equation}\label{set2}
 [\mathcal{Q}_0,\mathcal{Q}_a]=0,\quad[\mathcal{Q}_a,\mathcal{Q}_b]=2i\varepsilon_{abc}\mathcal{Q}_c, \quad a,b,c=1,2,3.
\end{equation}
The existence of the operators (\ref{set3}) is a direct consequence of the time-reversal symmetry of (\ref{totalh}). Indeed, (\ref{tr}) implies the unitary equivalence of the valley Hamiltonians $h(\pm\mathbf{K})$ and enables the construction of antidiagonal operators (\ref{set3}). 

The action of the integrals is quite nontrivial and determined by the properties of the Lax operator $S_{N+1}$. It can be inferred from (\ref{Skernel}) that all doublet states $\Psi_{2-deg}$ of  $\mathcal{H}$, corresponding to the band-edge energies $\lambda_n$, are annihilated by  the integrals of motion $ \mathcal{Q}_a $,
\begin{equation}
 \mathcal{Q}_a \Psi_{2-deg}=0, \quad a=0,1,2,3 \, .
\end{equation}
Let us denote by the subscript $_{\mathbf{K}}$ and  $_{\mathbf{K}'(=-\mathbf{K})}$ the states that are nonvanishing in one valley only, i.e. $\frac{1}{2}\left(1\pm\tau_{30}\right)\Psi_{\pm \mathbf{K}}=\Psi_{\pm \mathbf{K}}$.  We can find mutual eigenstates $\Psi^{\pm}_{\mathbf{K}}$ and $\Psi^{\pm}_{\mathbf{K'}}$ of the Hamiltonian $\mathcal{H}$, the valley-index operator $\tau_{30}$, and the integrals $\mathcal{Q}_0$ and $\mathcal{Q}_3$. They satisfy the following relations,
\begin{equation}
 (\mathcal{H}-\lambda)\Psi^{\pm}_{\mathbf{K}(\mathbf{K}')}=0,\quad (\tau_{30}-1)\Psi^{\pm}_{\mathbf{K}}=(\tau_{30}+1)\Psi^{\pm}_{\mathbf{K'}}=0
\end{equation}
 and
\begin{equation}\label{r1}
 \mathcal{Q}_i \Psi^{\pm}_{\mathbf{K}}  =  \pm\gamma_\lambda  {\Psi}^{\pm}_{\mathbf{K}}, \quad  \mathcal{Q}_i \Psi^{\pm}_{\mathbf{K'}}  =  \pm\gamma_\lambda  {\Psi}^{\pm}_{\mathbf{K'}}     \quad i=0,3 \, .
\end{equation}
The eigenvalues $\gamma_{\lambda}$ can be determined from the spectral polynomial (\ref{BuchanalChaudy}) as
\begin{equation}\label{gamma}
\gamma_\lambda = \sqrt{P(\lambda)}= \prod_{n=0}^{2N+1}(\lambda-\lambda_{n})^{1/2}\,.
\end{equation}
Hence, the operators $\mathcal{Q}_0$ and $\mathcal{Q}_3$ act on the basis of $\Psi^{\pm}_{\pm\mathbf{K}}$ as the multiplication by $\sqrt{P(\lambda)}$, i.e. as the square root of the operator $P(\mathcal{H})$. As mentioned above, the roots of the spectral polynomial (\ref{gamma}) coincide with the van Hove singularities of the analyzed finite-gap systems. 
The two antidiagonal operators $\mathcal{Q}_1$ and $\mathcal{Q}_2$ switch the valley index,
\begin{equation}\label{r2}
 \mathcal{Q}_1 \Psi^{\pm}_{\mathbf{K}(\mathbf{K}')}  =  \pm\gamma_\lambda  {\Psi}^{\pm}_{\mathbf{K}'(\mathbf{K})}, \quad  \mathcal{Q}_2 \Psi^{\epsilon}_{\pm\mathbf{K}}  =  \pm i\epsilon\gamma_\lambda  {\Psi}^{\epsilon}_{\mp\mathbf{K}} \, ,
\end{equation}
where $\epsilon=\pm$.

The action of the operators $\mathcal{Q}_a$ on the valley index is not indicated by the algebra (\ref{set2}). To reflect better the properties of the system, we can define a superalgebra graded by the valley index operator $\tau_{30}$. 
We denote $\mathcal{F}_{1(2)}\equiv \mathcal{Q}_{1(2)}$ as fermionic operators that change the valley index of the wave functions ($\{\mathcal{F}_{1(2)},\tau_{30}\}=0$) and $\mathcal{B}_{1(2)}\equiv\mathcal{Q}_{0(3)}$ as bosonic operators that preserve the valley index ($[\mathcal{B}_{1(2)},\tau_{30}]=0$).
The superalgebra is nonlinear and contains other fermionic operators $\tau_{12}$ and $\tau_{22}$,
\begin{eqnarray}\label{susy1}
 &&[\mathcal{H},\mathcal{B}_a]=[\mathcal{H},\mathcal{F}_a]=0,\ \{\mathcal{F}_a,\mathcal{F}_b\}=2\delta_{ab}P(\mathcal{H}),\\
 && [\mathcal{B}_a,\mathcal{F}_{b}]=2i\,\delta_{a2}\,\epsilon_{3bc}\,\tau_{c2}\,P(\mathcal{H}),\\
 && [\mathcal{B}_a,\tau_{22}]=-2i\delta_{2a}\,\mathcal{F}_{1},\quad [\mathcal{B}_a,\tau_{12}]=2i\,\delta_{2a}\,\mathcal{F}_{1},\\ 
&& \{\mathcal{F}_a,\tau_{b2}\}=2\delta_{ab}\,\mathcal{B}_1.\label{susy3}
\end{eqnarray}
The fact that we deal with finite-gap systems is manifested in the anticommutator of the fermionic operators where the spectral polynomial $P(\mathcal{H})$ emerges naturally.  It underlies nonlinearity of the superalgebra and manifests the intimate relationship of between the algebraic structure and the van Hove singularities of the considered models. 

Let us stress that the superalgebra (\ref{susy1})-(\ref{susy3}) exists for any finite-gap configuration of the twisted nanotubes described by $\mathcal{H}$ as long as the Hamiltonian possesses the time-reversal symmetry. 
 
The choice of the grading operator was not unique.  We could use either $\tau_{12}$ or $\tau_{22}$ equally well; both of them either commute or anticommute with the considered operators. Notice that $\tau_{12}$ corresponds to the unitary component of the time-reversal (\ref{tr}).
Choosing any of them as the new grading operator, qualitatively the same superalgebra would be obtained. The operators (\ref{q03})-(\ref{set3}) would be just permuted in the role of bosonic and fermionic generators.

Let us notice that in examples  presented in the previous section,  the single valley Hamiltonians with the vector potentials (\ref{crystal1}) and (\ref{crystal4}) commute with the operator $\sigma_3\mathcal{R}$ where $\mathcal{R}$ is the parity\footnote{ The three-gap setting with (\ref{crystal2}) has the nonlocal integral of slightly modified form, see \cite{BdG}.}, $\mathcal{R}x\mathcal{R}=-x$.
Hence, the corresponding Hamiltonian (\ref{totalh}) is commuting with $\tau_{33}\mathcal{R}$. The latter operator also commutes with $\tau_{30}$ and  $\tau_{22}$, whereas it anticommutes with all the operators $\mathcal{Q}_a$, $a=0,..,3$. It means that $\tau_{33}\mathcal{R}$ could be regarded as a grading operator of a distinct, $N=4$ superalgebra that would be generated by four fermionic operators (\ref{q03}) and (\ref{set3}). The nonlinear superalgebra of Bogoliubov-de Gennes Hamiltonians generated by nonlocal supercharges was discussed in the literature. We  refer to \cite{BdG} for more details, see also \cite{varios papers}. 

The formulas for LDOS and DOS computed in the third section with the use of the formula (\ref{poly}) have to be multiplied by four to get the correct form for the corresponding twisted nanotubes. Indeed, we have to take into account the valley degeneracy that we discussed in this section, as well the double degeneracy of energy levels due to (real) spin$-\frac{1}{2}$ of the particle that was neglected up this moment.

Finally, let us discuss briefly the settings where an external magnetic field is present in addition to the twists. The magnetic field breaks the time-reversal symmetry. When the vector potential $\Delta_{mg}$ is included into the Hamiltonian, we have 
\begin{equation}\label{totalhmg}
 h(\pm\mathbf{K})=-i\partial_{x}\sigma_2\pm \Sigma(x) \sigma_1\pm M\sigma_3+\Delta_{mg}\sigma_1.
\end{equation}
We can see that as long as mass term $M$ is vanishing and either magnetic field \textit{or} twists are switched on (i.e. $\Sigma(x)\Delta_{mg}=0$), all the energy levels have even degeneracy due to the unitary equivalence of $h(\mathbf{K})$ and $h(\mathbf{K'})$. The situation changes when both $\Sigma(x)$ and $\Delta_{mg}$ are nonzero. In that case, we can still have a finite-gap configuration in one valley described by $h(\mathbf{K})$. However, in the second valley the finite-gap potential is violated in general by the changed sign of $\Delta_{mg}$. 

Curiously enough, we can still get a finite-gap configuration for each valley by the fine-tuning of the external field. 
As an example, let us consider the situation when the low-energy dynamics in the $\mathbf{K}$ valley is described by 
\begin{equation}\label{twistmg1}
 h(\mathbf{K})=-i\partial_x\sigma_2+(\coth b+\tanh x-\tanh (x+b))\sigma_1,
\end{equation}
which is an infinite-period limit of the three-gap system (\ref{crystal2}). Let us suppose that the vector potential in (\ref{twistmg1}) is induced both by radial twist and by external magnetic field,  where
$\Sigma(x)=\frac{1}{2}\left(2\coth b+\tanh x-\tanh (x+b)\right)$ and $\Delta_{mg}=\frac{1}{2}\left(\tanh x-\tanh (x+b)\right)$. Then the subsystem in the $\mathbf{K}'$-valley is described by
\begin{equation}\label{twistmg2}
 h(\mathbf{K'})=-i\partial_x\sigma_2-\coth b\, \sigma_1,
\end{equation}
which is just the trivial one-gap system. In the current setting, deformation associated with $\Sigma(x)$ is asymptotically uniform but gets changed in the localized region where the (asymptotically vanishing) external magnetic field is nonzero.  The spectrum of the corresponding total Hamiltonian $\mathcal{H}$  has two singlet discrete energy levels corresponding to the bound states and two doubly degenerate levels $\pm \coth\, b$ corresponding to the threshold of the positive and negative continuum. The other energy levels are four-fold degenerate.

It is worth noticing that the discussed framework can be understood in the context of (planar) graphene crystal in the presence of the external magnetic field and strain, both of which depend on $x$ only. Due to separability of the stationary equation, the one-dimensional Hamiltonian can be written as 
\begin{equation}\label{plane}
 h(\mathbf{K})=-i\sigma_2\partial_{x}+(k_y+A_y(x))\sigma_1,
\end{equation} 
where $k_y$ corresponds to the momentum that is parallel with the (pseudo-) magnetic barrier. The operator (\ref{plane}) describes a massless Dirac particle that moves with fixed direction in the presence of vector potential $A_y$, associated with the strain and the external magnetic field. 
In this context, the setting with the single-valley Hamiltonians (\ref{twistmg1}) and (\ref{twistmg2}) with the inhomogeneous external magnetic field perpendicular to the surface and given by $\Delta_{mg}$ is rather realistic. 

\section{Discussion and Outlook}\label{conclusion}
The one-dimensional Dirac operator with finite-gap potential appears in a variety of physically interesting models \cite{Dunne}, \cite{BdG}, \cite{Feinberg}, \cite{PlyushchayBdG}. In the present paper, we illustrated how the machinery of the AKNS hierarchy can be used in the analysis of the twisted nanotubes in the low-energy regime,  particularly, for the computation of the local density of states.

We showed that the  finite-gap, time-reversal invariant configurations possess a hidden nonlinear supersymmetry  that is associated with the Kramer's degeneracy of energy levels. Physics of  these systems, namely the presence of the two valleys at  $\mathbf{K}$ and $\mathbf{K'}$ and the preserved time-reversal symmetry, is responsible for the form of the Hamiltonian (\ref{totalh}) which consists of two copies of the (unitarily) equivalent  single-valley energy operators.

The current situation differs from the quantum models with bosonized supersymmetry \cite{bosonized}, where  a nonlocal integral of motion was identified as the grading operator. 
Both the Hamiltonian (\ref{totalh}) and its integrals of motion (\ref{q03}) and (\ref{set3}), forming  $so(3)\oplus u(1)$ Lie algebra, can be graded by a local operator, e.g. by the valley index operator $\tau_{30}$. This framework represents a nontrivial example of the hidden supersymmetry in the sense that it naturally emerges within the unextended, physical Hamiltonian (\ref{totalh}). 

The explicit results for the presented finite-gap systems can be extended with the use of Darboux transformation \cite{DiracDarboux}.  Within this framework, one can construct new finite-gap Hamiltonians $h_2$ from a known  one, namely $h_1$.  The transformation is given in terms of a matrix differential operator $D$, which intertwines two one-dimensional Dirac Hamiltonians, $Dh_1=h_2D$. It maps the eigenstates of $h_1$ into those of $h_2$, keeping the operators (almost) isospectral. Moreover, the diagonal resolvent of $h_2$ can be computed directly from the diagonal resolvent of $h_1$ with the use of $D$, see \cite{twisting} for details.

In the paper, the operator (\ref{eq1m}) was almost exclusively interpreted as the effective Hamiltonian of the twisted carbon (or boron-nitride) nanotube. As we discussed in the end of the preceding section, the results can be also used in the analysis of the Dirac electrons in graphene in the presence of (pseudo-)magnetic barriers.
Such systems with Kronig-Penney or a piece-wise constant (pseudo-) magnetic fields induced by either external field or strains were considered in the literature, see \cite{KP} or \cite{steplikeB}. 
In this context, the DOS computed for the twisted nanotubes can be interpreted as the partial density of states in graphene for the $k_y=0$ channel. It could facilitate the computation of the transition coefficient in the normal direction to the magnetic barrier. The known results \cite{partialLDOS} on the relation between one-dimensional DOS and the phase of the transmission amplitude could be particularly helpful in this context.

Considering Dirac electrons in graphene, it is desirable to extend the analysis for $k_y\neq 0$ as well. Keeping in mind \cite{milpas} or \cite{Miserev}, the infinite-period limit of the finite-gap models could be a feasible starting point in this respect. The analysis of periodic systems could make it possible to observe the phenomena that appear in graphene superlattices, the new generation of Dirac points in particular  \cite{newgeneration}, \cite{steplikeB2}.

Study of the finite-gap configurations of electrostatic potential represents another possible direction for future research. Spectral properties of Dirac electron in graphene in presence of both periodic electrostatic and magnetic fields were discussed in various works, see e.g. \cite{Novikov}, \cite{Dugaev}, \cite{Wang}. In this context, the mapping between the systems with magnetic and electrostatic field \cite{tan} could provide an interesting way to extend our results. 

The finite-gap systems are an approximation of more realistic settings. They can serve as a test field for numerical or perturbative methods and can also provide qualitative insight into the experimental data. 
Although, to our best knowledge,  the experiments with the single-wall carbon nanotubes with the periodically modulated twist have not been prepared yet, the building blocks of such settings seem to be available, see e.g. \cite{pendulum}-\cite{rotors}. 

As an experimental implementation of the proposed models, we can imagine a long suspended nanotube anchored to a substrate at the ends and a periodic array of small paddles attached to it. By deflection of the paddles, the twist of the nanotube could be altered. Let us mention that the latter configuration (with one paddle) was employed in \cite{torsionalprops} for measurement of the torsional properties of the nanotubes. 

In the presented finite-gap models, the possible interaction of the nanotube with the anchors and the paddles is not taken into account. Still, we think that the results (e.g. the suggested dependence of the central gap on the average twist) might provide interesting qualitative insight into the spectral properties of the settings realized in the experiments.

\acknowledgments 
The authors would like to thank to Gerald Dunne and Dmitry V. Kolesnikov for discussions.
F.C. was supported by the Fondecyt Grant No. 11121651 and by the Conicyt grant  79112034 and ACT-91.  
F.C. wishes to thank the kind hospitality of the Nuclear Physics Institute of the ASCR and the Physics
Department of the University of Connecticut. The
Centro de Estudios Cient\'{\i}ficos (CECs) is funded by the Chilean Government
through the Centers of Excellence Base Financing Program of Conicyt. V.J. was supported by  GA\v CR Grant No. P203/11/P038 of the Czech Republic.

\end{document}